\documentclass[confenrence]{IEEEtran}
\usepackage{amsmath,amsfonts}
\usepackage{hyperref}
\usepackage{algorithmic}
\usepackage{array}
\usepackage[caption=false,font=normalsize,labelfont=sf,textfont=sf]{subfig}
\usepackage{textcomp}
\usepackage{stfloats}
\usepackage{url}
\usepackage{verbatim}
\usepackage{graphicx}
\def\BibTeX{{\rm B\kern-.05em{\sc i\kern-.025em b}\kern-.08em
    T\kern-.1667em\lower.7ex\hbox{E}\kern-.125emX}}
\usepackage{balance}

\begin{document}

\title{Microwave SQUID Multiplexer Readout Performance Using a Direct-RF RFSoC-Based Software-Defined Radio}

\author{\IEEEauthorblockN{M. E. Garc\'ia Redondo\IEEEauthorrefmark{1}\IEEEauthorrefmark{2}\IEEEauthorrefmark{5}\IEEEauthorrefmark{6},
J. D. Bonilla Neira\IEEEauthorrefmark{1}\IEEEauthorrefmark{3}\IEEEauthorrefmark{4}\IEEEauthorrefmark{6},
N. A. Müller\IEEEauthorrefmark{1}\IEEEauthorrefmark{3}\IEEEauthorrefmark{5}\IEEEauthorrefmark{6}, L. P.  Ferreyro\IEEEauthorrefmark{1}\IEEEauthorrefmark{2}\IEEEauthorrefmark{4}\IEEEauthorrefmark{6}, J. M. Geria\IEEEauthorrefmark{1}\IEEEauthorrefmark{3}\IEEEauthorrefmark{4}\IEEEauthorrefmark{6},\\ T. Muscheid\IEEEauthorrefmark{2}, R. Gartmann\IEEEauthorrefmark{2}, A. Almela\IEEEauthorrefmark{1}\IEEEauthorrefmark{5}, M. R. Hampel\IEEEauthorrefmark{1}\IEEEauthorrefmark{4}\IEEEauthorrefmark{5}\IEEEauthorrefmark{6}, L. Ardila-Perez\IEEEauthorrefmark{2}, M. Wegner\IEEEauthorrefmark{2}\IEEEauthorrefmark{3},\\ M. Platino\IEEEauthorrefmark{1}\IEEEauthorrefmark{4}\IEEEauthorrefmark{5}\IEEEauthorrefmark{6}, O. Sander\IEEEauthorrefmark{2}, S. Kempf\IEEEauthorrefmark{2}\IEEEauthorrefmark{3} and M. Weber\IEEEauthorrefmark{2}}

\IEEEauthorblockA{\IEEEauthorrefmark{1}Instituto de Tecnolog\'ias en Detección y Astropart\'iculas (ITeDA), Argentina\\
\IEEEauthorrefmark{2}Institute for Data Processing and Electronics (IPE), Karlsruhe Institute of Technology (KIT), Germany\\
\IEEEauthorrefmark{3}Institute of Micro-and Nanoelectronic Systems (IMS), Karlsruhe Institute of Technology (KIT), Germany\\
\IEEEauthorrefmark{4}Consejo Nacional de Investigaciones Cient\'ificas y T\'ecnicas (CONICET), Argentina\\
\IEEEauthorrefmark{5}Comisi\'on Nacional de Energ\'ia At\'omica (CNEA), Argentina\\
\IEEEauthorrefmark{6}Universidad Nacional de San Mart\'in (UNSAM), Argentina}}


\maketitle

\begin{abstract}
In this work, we report the experimental readout results of a \textmu MUX device using a Direct-RF SDR prototype based on the ZCU216 Radio-Frequency System-on-Chip (RFSoC) evaluation board. First, the analog performance of the SDR system was evaluated both in loopback and coupled to the cryogenic multiplexing system. Then, the SDR system performance was optimized for \textmu MUX readout, focusing on bolometric applications. Finally, the minimum demodulated flux noise was obtained through the optimally conditioned readout of a \textmu MUX channel. The results presented are comparable to those obtained with traditional SDR architectures and demonstrate that the Direct-RF SDR prototype is suitable for the readout of \textmu MUX devices. We anticipate that Direct-RF SDR technology will play a key role in enabling the next generation of SDR readout systems for frequency-multiplexed low-temperature detector arrays.
\end{abstract}


\section{Introduction}

\noindent \IEEEPARstart{T}{he} Microwave SQUID Multiplexer (\textmu MUX) has become widely used for the multiplexed readout of low-temperature detector arrays, as it enables high multiplexing factors while keeping the readout noise well below the detector noise~\cite{Bennett2019,Dober2017}. The \textmu MUX encodes detector signals in the resonance frequencies of multiple GHz-frequency superconducting resonators, which can be monitored externally from the cryostat using a Software-Defined Radio (SDR) system. As a result, the cryogenic multiplexing complexity is reduced and shifted to room temperature, where it can be managed with modern SDR technology.

Traditionally, SDR readout systems have relied on high-speed Digital-to-Analog Converters (DACs) and Analog-to-Digital Converters (ADCs) in combination with mixing boards to access the \textmu MUX frequency band~\cite{Yu2023-dk,Muscheid2024}. While these schemes provide straightforward frequency conversion and demodulation of complex signals, they suffer from the drawbacks associated with analog mixers. These include IQ imbalance and LO leakage, which require a tedious and frequent calibration process~\cite{herrmann2022} Nowadays, the most advanced commercially available Direct-RF sampling SDRs offer a solution to this problem. The high sampling rates of these devices enable direct access to the microwave frequency band~\cite{rfsoc,agilex,versal}. At the same time, they are equipped with Digital-Up Converters (DUCs) and Digital-Down Converters (DDCs), allowing digital implementation of IQ sampling without being affected by the issues inherent to analog mixers. While this SDR technology has been driven by several recent developments~\cite{Sinclair2022-wz,Stefanazzi2022,liu2024} and represents a compact and flexible solution, there is still insufficient evidence demonstrating that the analog performance of these systems is suitable for \textmu MUX readout. 

This work reports the readout noise performance results of a Direct-RF Software-Defined Radio (SDR) system designed for the readout of low-temperature detectors using a \textmu MUX.

\section{Direct-RF RFSoC-Based SDR Architecture}

\begin{figure*}[t]
\centering
    \centering
    \captionsetup[subfigure]{
        singlelinecheck = false,
        justification=raggedright,
        captionskip=0mm,
        position=top
    }
    \subfloat[\label{fig:directrf-sdr}]{{\includegraphics[width=0.48\textwidth]{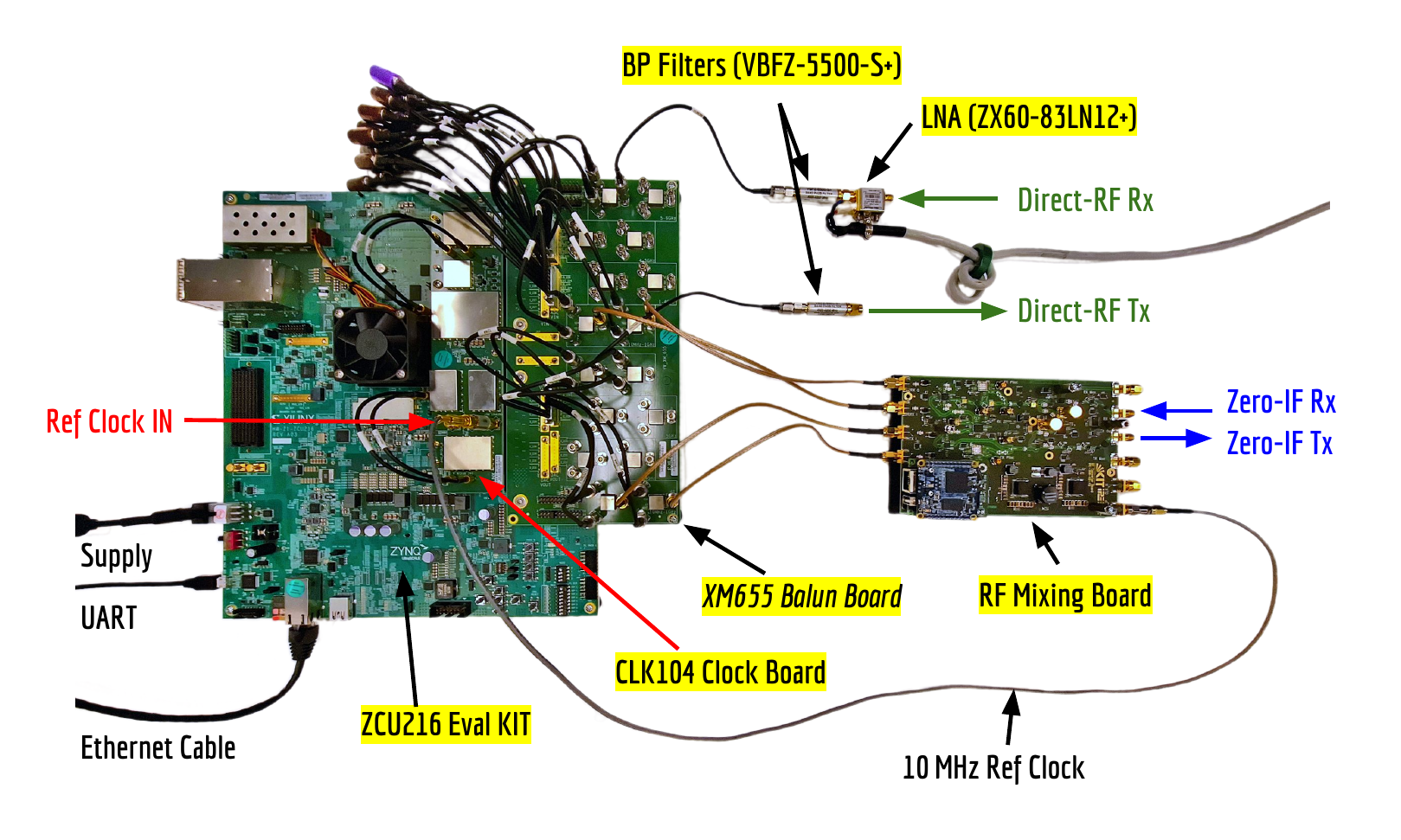} }}%
    \subfloat[\label{fig:directrf-firmware}]{{\includegraphics[width=0.48\textwidth]{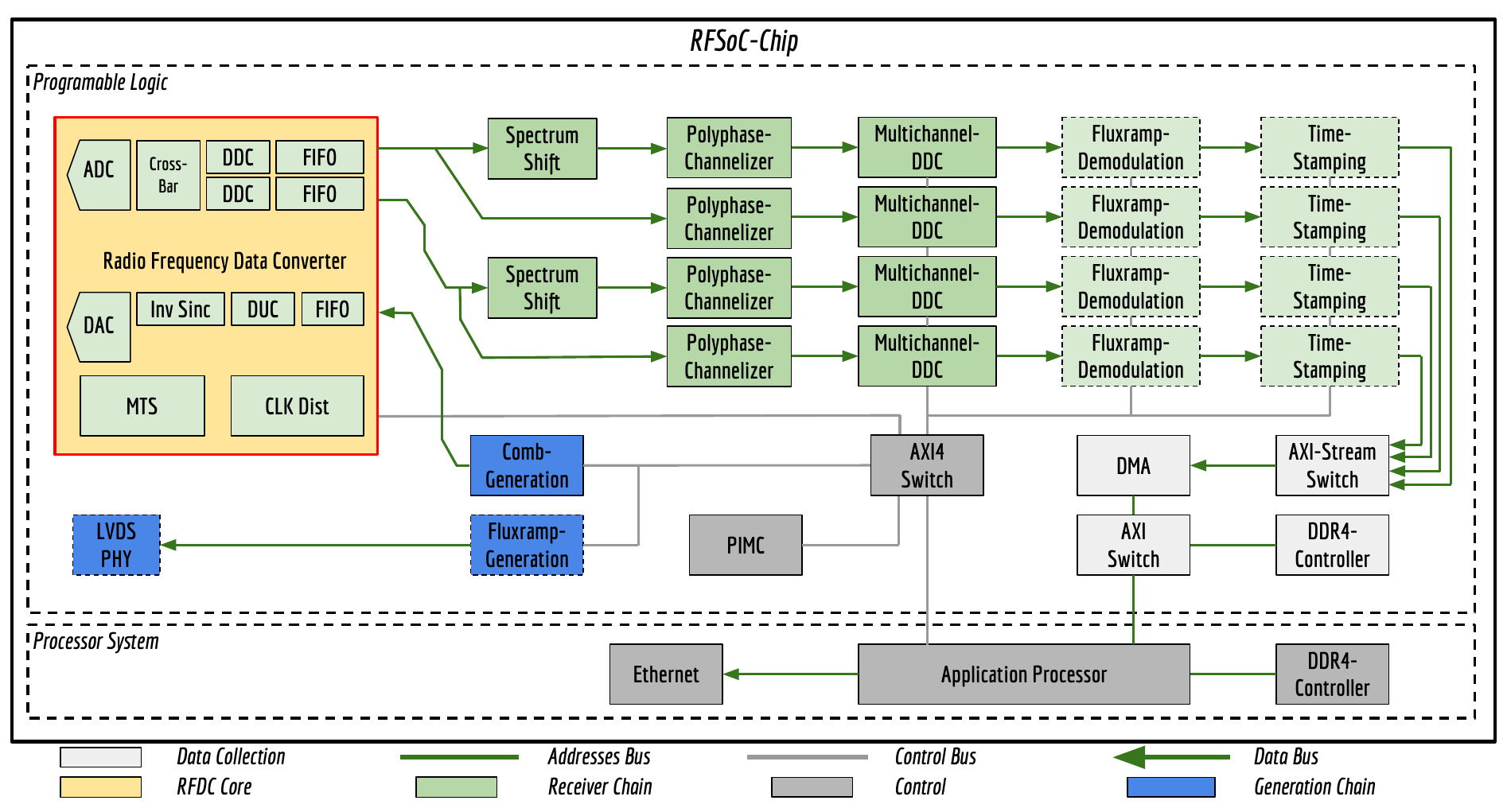} }}%
    \label{fig:example}
\caption{(a) Hardware implementation of the Direct-RF RFSoC-Based SDR prototype systems implemented using the ZCU216 evaluation board. The Direct-RF SDR system, shares the evaluation kit with the previously developed Zero-IF SDR system~\cite{Redondo2024}. (b) Block diagram of the SDR firmware implemented in the RFSoC chip, inherited from previous developments. The block highlighted in red corresponds to the Radio-Frequency Data Converters IP core~\cite{rfdc}, which was configured to implement the Direct-RF SDR architecture.}
\end{figure*}

\noindent The Direct-RF RFSoC-based SDR system, illustrated in Figure~\ref{fig:directrf-sdr}, is implemented on the ZCU216 evaluation kit~\cite{zcu216}, which is also employed in the previously developed Zero-IF SDR system~\cite{Redondo2024,Gartmann2024}. The evaluation kit is complemented with the \textit{XM650} BalUn board for differential-to-single-ended conversion and the \textit{CLK104} board for sampling clock generation~\cite{clk104}. Furthermore, two \textit{Mini-Circuits VBFZ-5550-S+} band-pass filters and a wideband low-noise amplifier (LNA), the \textit{Mini-Circuits ZX60-83LN12+}, are incorporated to provide gain and suppress unwanted image tones in transmission/reception. The transmitter and receiver ports are represented with green arrows in Figure~\ref{fig:directrf-sdr}. The firmware architecture shown in Figure~\ref{fig:directrf-firmware} was adopted from previous developments~\cite{Muscheid2024,Redondo2024}. It is responsible for generating and acquiring the frequency comb as well as for performing fine and coarse channelization as well demodulation. The block highlighted in red corresponds to the Radio-Frequency Data Converters (RFDC) IP core~\cite{rfdc} which was configured to implement the Direct-RF SDR. The RFDC configuration described in the next sections preserves the input/output data rates, thereby ensuring compatibility.

\subsection{Direct-RF Signal Generation}

\begin{figure}[!b]
\centering
    \centering
    \captionsetup[subfigure]{
        singlelinecheck = false,
        justification=raggedright,
        captionskip=0mm,
        position=top
    }
    \subfloat[\label{fig:directrf-DAC}]{{\includegraphics[width=0.48\textwidth]{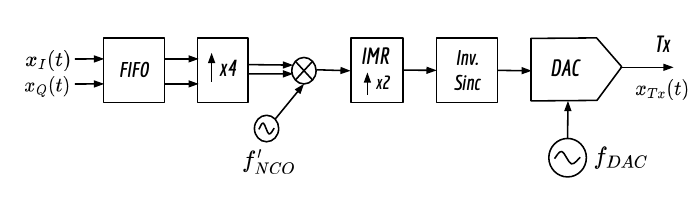} }}
    \qquad
    \subfloat[\label{fig:directrf-ADC}]{{\includegraphics[width=0.48\textwidth]{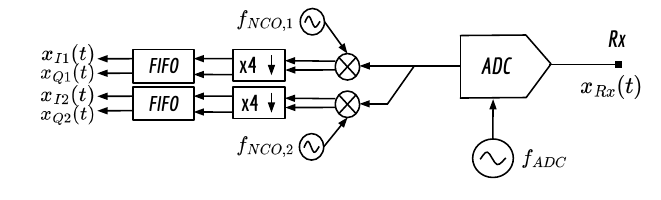} }}%
    \caption{(a) Direct-RF SDR RFDC DAC configuration. The digital data-path is configured as Complex-to-Real and the DAC is sampling at $f_{DAC}=8$~GHz. (b) Direct-RF SDR RFDC ADC configuration. The digital data-path is configured as Real-to-Complex and the ADC is sampling at $f_{ADC}=2$~GHz.}
\end{figure}

\noindent The third-generation RFSoC DACs support sampling rates up to $f_{DAC}=9.85$~GHz~\cite{rfsoc} and include different reconstruction waveforms~\cite{rfdc}. In the proposed Direct-RF SDR, the DACs use the mix-mode reconstruction waveform and operate at $f_{DAC}=8$~GHz generating a single $800$~MHz band within $4$–$8$~GHz. The DAC digital data path configuration is shown in Figure \ref{fig:directrf-DAC}. Here, the comb generator drives the DAC through a 256-bit complex AXI4-Stream at $f_{DAC}^{AXI}=125$~MHz, carrying sixteen words per clock. Complex data stream at $f_{s}=1$~GHz is interpolated by $\times 4$ before entering the DUC in IQ-to-Real mode, where it is up-converted by an NCO at $f_{NCO}^{\prime}$. Since the DUC is limited to $6$~GHz, an additional $\times 2$ interpolation is provided by the Image Rejection Filter (IMR) configurable as low-pass or high-pass~\cite{rfdc}. Therefore, the effective NCO frequency is $f_{NCO}=2 f_{NCO}^{\prime}$. It is important to note that the operation in the second Nyquist zone introduces a spectrum mirroring that was compensated by choosing $-f_{NCO}$.

\subsection{High-Order Nyquist Zone Signal Acquisition}

\noindent The ZCU216 kit integrates a ZU49DR device~\cite{rfsoc}, which provides four ADCs per tile with a maximum sampling rate of $f_s=2.5$~GHz~\cite{rfdc}. Consequently, frequency comb acquisition above $4$~GHz relies on band-pass sampling in higher Nyquist zones. As can be seen in Figure~\ref{fig:directrf-ADC}, the received signal $x_{Rx}(t)$ is sampled at $f_{ADC}=2$~GHz, folding signals from higher Nyquist zones onto the first one. Therefore, each zone has a maximum bandwidth of $f_s/2=1$~GHz. As a consequence, the DAC-generated bands by design were centered at $f_{NCO}=4.5+1n$~GHz ($n\in[0,3]$) can be acquired after filtering. Although the absence of analog mixers is advantageous, it requires sharp microwave filters and guard bands, limiting the implementation to four bands. After signal acquisition, the ADC output is split into two DDCs in Real-to-Complex mode, with NCOs at $f_{NCO,1}=296.875$~MHz and $f_{NCO,2}=703.125$~MHz to separate upper and lower sub-bands centered at $f_{c}=500$~MHz. A $\times 4$ decimator reduces the rate, and two 128-bit complex AXI streams (i.e. each one with 8 words per clock at $f_{ADC}^{AXI}=125$~MHz) deliver the data to the logic. Since even zones fold with spectral inversion, this can corrected using $-f_{NCO,-1}$ and $-f_{NCO,2}$.

\section{SDR Readout System Characterization}

\noindent The following subsections provides a detailed description of the SDR systems characterization steps. 

\subsection{Transmission Characterization}

\begin{figure}[!t]
\centering
    \centering
    \captionsetup[subfigure]{
        singlelinecheck = false,
        justification=raggedright,
        captionskip=0mm,
        position=top
    }
    \subfloat[\label{fig:setup_tx}]{{\includegraphics[width=0.48\textwidth]{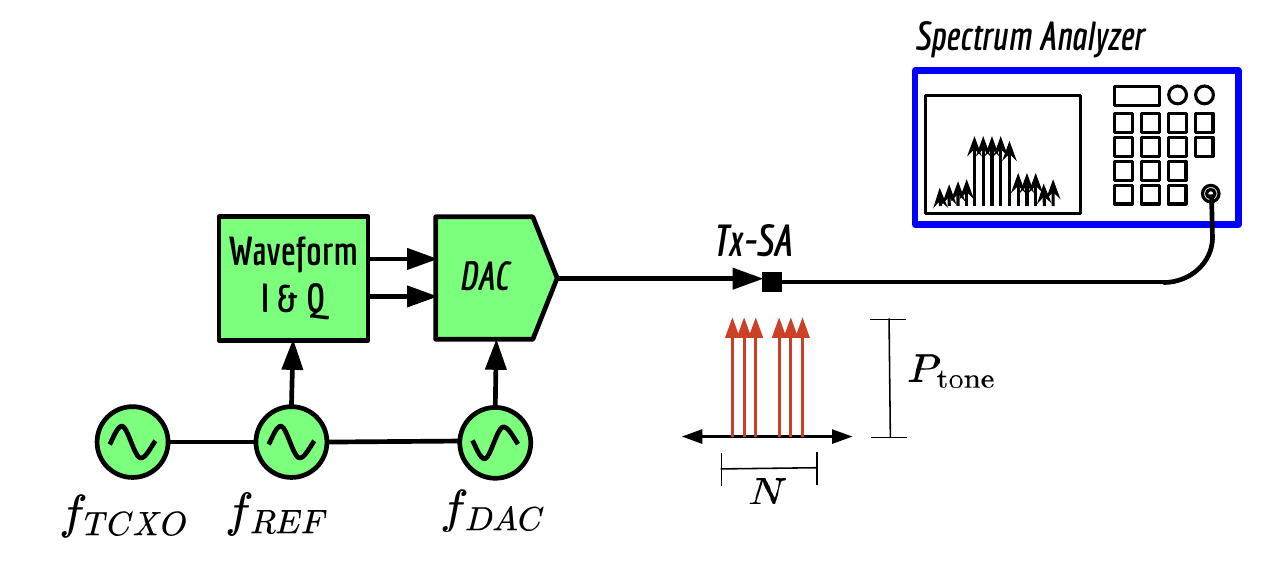} }}%
    \qquad
    \subfloat[\label{fig:setup_loop}]{{\includegraphics[width=0.48\textwidth]{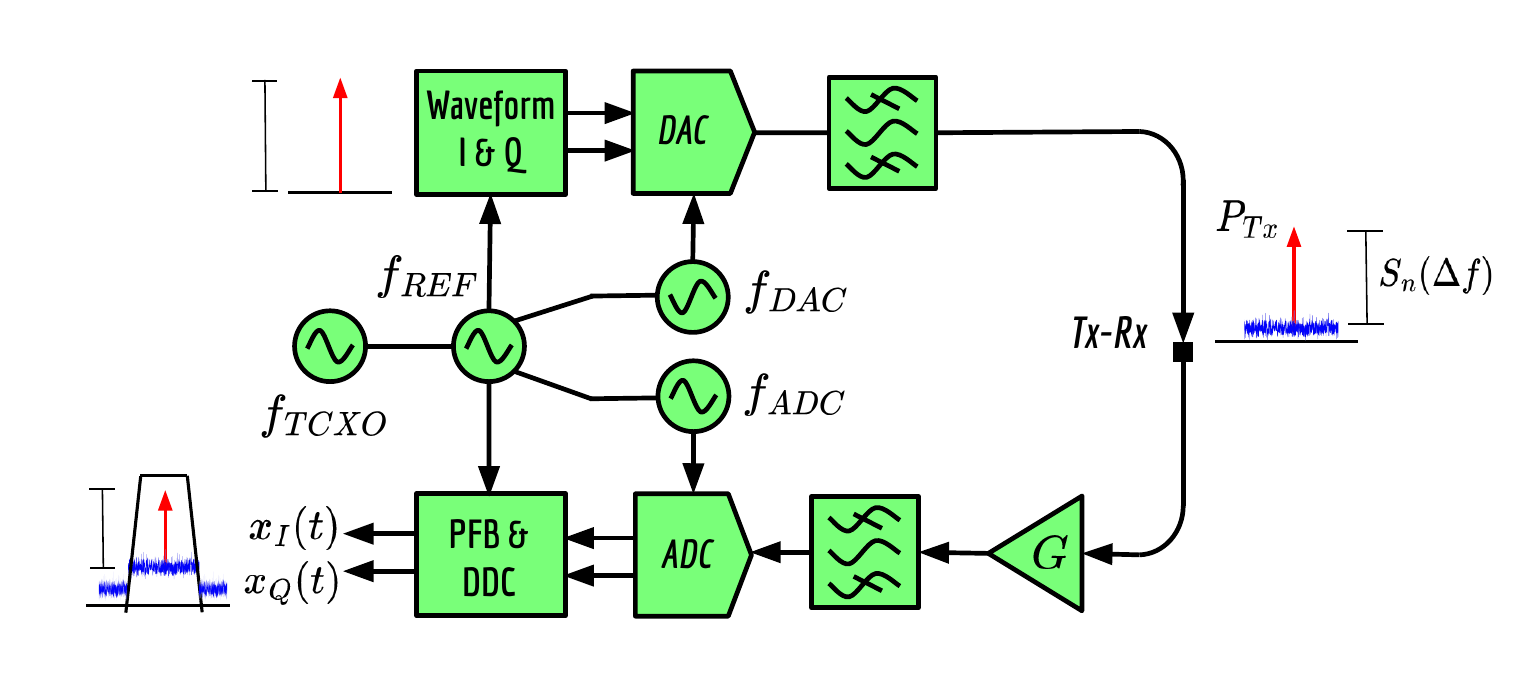} }}%
\caption{(a) Block diagram of the Direct-RF SDR Characterization Set-Up in transmission using an spectrum analyzer. (b) Block diagram of the Direct-RF SDR Characterization Set-Up in RF-Loopback.}
\end{figure}

\noindent The transmission characterization process aimed to determine the system's ability to generate a frequency comb, as well as the gain and filtering requirements necessary to achieve the signal quality required by the \textmu MUX. For this case, $N=100$ tones were generated, spanning a total bandwidth of $\approx 800$~MHz and reproduced at a rate of $f_{DAC}=8$~GHz.This value corresponds to the maximum allowed by the interpolation filter response of the RFDC~\cite{rfdc} for the given sampling rate. To cover the entire frequency range between $4$ and $8$~GHz, the tones were sequentially centered at $f_{NCO}=4.5 + 1n$~GHz with $n \in [0,3]$ and measured using an \textit{R\&S FSWP 50} spectrum analyzer. The measurement results are shown in Figure~\ref{fig:setup_tx}. Since the \textit{XM650} Balun board does not provide the full frequency range in single SMA connector, the DAC output was connected to the appropriate Balun for each $f_{NCO}$ frequency. Specifically, the \textit{Anaren BD3150N50100AHF} and \textit{Anaren BD4859N50100AHF} Baluns were employed for the $4$–$5$~GHz and $5$–$8$~GHz bands, respectively.

\begin{figure}[!t]
\centering
    \centering
    \captionsetup[subfigure]{
        singlelinecheck = false,
        justification=raggedright,
        captionskip=0mm,
        position=top
    }
    \subfloat[\label{fig:tones_tx}]{{\includegraphics[width=0.48\textwidth]{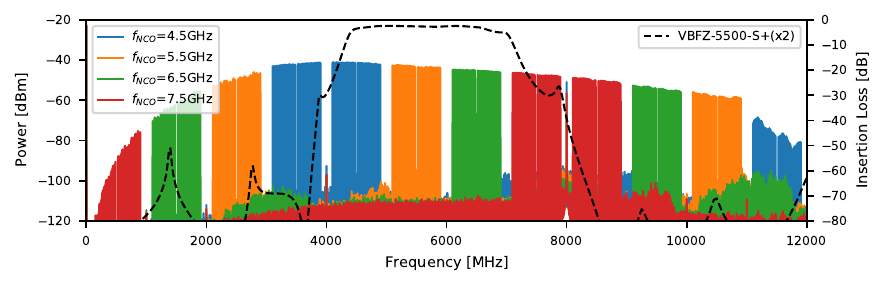} }}%
    \qquad
    \subfloat[\label{fig:tones_rx}]{{\includegraphics[width=0.48\textwidth]{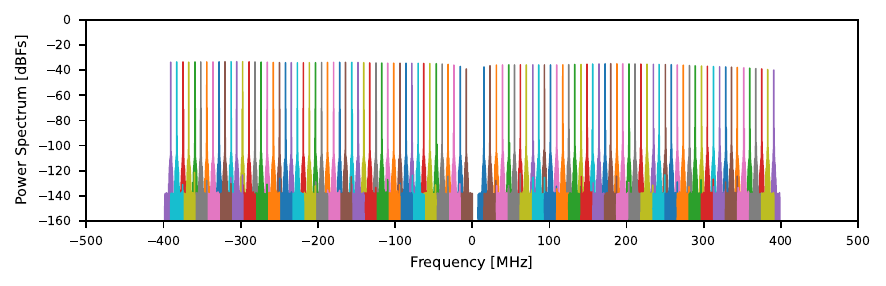} }}%
\caption{(a) Measured spectrum of $N=100$ tones generated by the Direct-RF SDR. Different colors represent four different NCO frequencies $f_{NCO}$. The DAC sampling frequency is $f_{DAC}=8$~GHz. The right vertical axis corresponds to the insertion loss response of the Band-Pass filter used for the characterization in RF-Loopback and is represented with a dashed black line. (b) $N=100$ tones centered at $f_{NCO}=6.5$~GHz acquired and channelized by the Direct-RF SDR prototype. The tones were generated by the Direct-RF SDR at a constant power of $P_{tone}=-45$~dBm.}
\end{figure}

\noindent Figure~\ref{fig:tones_tx} shows the spectrum of the tones generated by the Direct-RF SDR prototype using a total power of $P_{out} \equiv -5$~dBFS and measured with a spectrum analyzer over the bandwidth between $0$ and $12$~GHz. A previous single-tone characterization of the converters demonstrated that the use of the High Linearity (HL) decoder~\cite{rfdc} provides the best SFDR, approximately $55$~dB within the operating band between $4$ and $8$~GHz.  

From the results, it can be observed that no tones are present at the center of the band, while a series of high-power image tones appear on both sides of the operating band. The absence of LO leakage and IQ imbalance~\cite{herrmann2022} is due to the use of digital mixers. On the other hand, the appearance of spectral replicas originates from the reconstruction process. The DAC output signal contains all replicas of the analytic signal, weighted by the frequency response of the applied reconstruction waveform~\cite{Kalfus2020}. Since the signal is sampled at $f_{DAC} = 8$~GHz, each Nyquist zone spans $4$~GHz, and each replica is mirrored in the odd zones. This leads to one of the most significant drawbacks of this scheme: the tones at both edges of a Nyquist zone have replicas close in frequency. Therefore, the use of guard bands between sub-bands and selective band-pass filters should be considered in order to provide transition margins for the filters. A viable technological solution to this problem is the implementation of cavity filters, as reported in the following references~\cite{Yu2023-dk,liu2024}. Additionally, the measured power resulted in a slope of approximately $-8$~dB between the edges of the operating band. This behavior will be compensated in future work by using equalizers and amplifiers.

\subsection{RF-Loopback Characterization}\label{subsec:rfloop}

\noindent In order to evaluate channelization capabilities as well as noise performance, the transmitter and receiver were connected in an RF-Loopback configuration, as shown in Figure~\ref{fig:setup_loop}. During this measurement, two \textit{Mini-Circuits VBFZ-5550-S+} band-pass filters were used to remove unwanted images. Given the pass-band characteristics of the available filters (see dashed black line in Figure~\ref{fig:setup_loop}), only a single band centered at $f_{NCO}=6.5$~GHz was characterized. Figure~\ref{fig:tones_rx} demonstrates the capability of the SDR system to acquire and channelize a comb of $N=100$ tones.

To characterize the noise, the procedure previously described was implemented~\cite{Garcia2024}. A single tone with power $P_{Tx}$ at the transmitter (Tx) port was generated. It was then acquired, channelized, and down-converted to baseband using a Digital Down-Converter (DDC). From the IQ trace $x(t)=x_I(t)+j\,x_Q(t)$, an estimation of the Noise Spectral Density (NSD) was obtained for both phase $S_{\phi}^{\mathrm{meas}}(\Delta f)$ and amplitude $S_{\gamma}^{\mathrm{meas}}(\Delta f)$. In this case, $N=100$ NSD traces with $2^{25}$ samples were averaged. As expressed in Equations~\ref{eq:measphase} and \ref{eq:measamp}, the contributions of multiplicative noise (left term) and additive noise (right term) are measured simultaneously and cannot be distinguished. Here, $T_n^{Tx-Rx}$ denotes the sum of the additive noise temperatures of the transmitter $T_n^{Tx}$ and the receiver $T_n^{Rx}$.

\begin{equation}
    S_{\phi}^{\mathrm{meas}}(\Delta f)= S_{\phi}(\Delta f) +\frac{k_B T_{n}^{Tx-Rx}}{P_{Tx}},
    \label{eq:measphase}
\end{equation}

\begin{equation}
    S_{\gamma}^{\mathrm{meas}}(\Delta f)= S_{\gamma}(\Delta f) + \frac{k_B T_{n}^{Tx-Rx}}{P_{Tx}}.
    \label{eq:measamp}
\end{equation}

\begin{figure*}[t]
\centering
    \centering
    \captionsetup[subfigure]{
        singlelinecheck = false,
        justification=raggedright,
        captionskip=0mm,
        position=top
    }
    \subfloat[\label{fig:nsd_rf}]{{\includegraphics[width=0.47\textwidth]{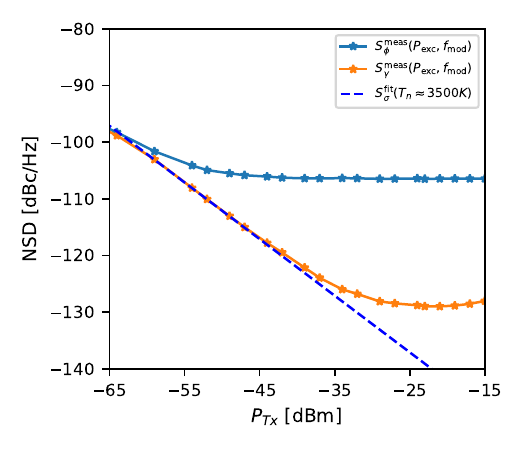} }}
    \subfloat[\label{fig:nsd_cryo}]{{\includegraphics[width=0.47\textwidth]{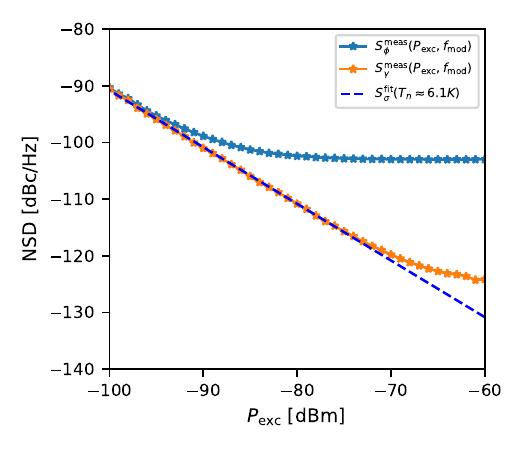} }}%
    \label{fig:example}
\caption{(a) Phase $S_{\phi}^{\mathrm{meas}}(P_{\mathrm{exc}},f_{\mathrm{mod}})$ and amplitude $S_{\gamma}^{\mathrm{meas}}(P_{\mathrm{exc}},f_{\mathrm{mod}})$ noise as a function of the transmitted power $P_{Tx}$ for a tone at $f_{\mathrm{exc}} \approx 6.4$~GHz measured in RF-Loopback. (b) Off-resonance phase $S_{\phi}^{\mathrm{meas}}(P_{\mathrm{exc}},f_{\mathrm{mod}})$ and amplitude $S_{\gamma}^{\mathrm{meas}}(P_{\mathrm{exc}},f_{\mathrm{mod}})$ noise measured in cryogenic loopback configuration as a function of the \textmu MUX excitation power $P_{\mathrm{exc}}$. Dashed Lines correspond to the fitted additive noise contribution $S_{\sigma}^{\mathrm{fit}}(P_{\mathrm{exc}},f_{\mathrm{mod}})$ as described in the main manuscript.}
\label{fig2}
\end{figure*}

The different contributions were separated by sweeping the transmitted power $P_{Tx}$ in the range $[-65,-15]$~dBm and fitting the measured phase $S_{\phi}^{\mathrm{meas}}(P_{\mathrm{exc}},f_{\mathrm{mod}})$ and amplitude $S_{\gamma}^{\mathrm{meas}}(P_{\mathrm{exc}},f_{\mathrm{mod}})$ noise densities as a function of the transmitted power $P_{Tx}$. Since the target applications are bolometric~\cite{Dober2021}, the fitting was restricted to noise values averaged over the possible flux-ramp modulation frequencies $f_{\mathrm{mod}}$ in the range $[5,50]$~kHz~\cite{Mates2012}. The measurement results are shown in Figure~\ref{fig:nsd_rf}, together with the fit of the additive component $S_{\sigma}^{\mathrm{fit}}(P_{\mathrm{exc}},f_{\mathrm{mod}})$, which is only appreciable at low powers. The values of multiplicative and additive noise are listed in Table~\ref{tab:noiselev_rf}. 

\begin{table}[h]
   \caption{Summary of the additive and multiplicative noise levels for the Direct-RF RFSoC-Based SDR prototype measured in RF-Loopback.}
   \label{tab:noiselev_rf}
    \centering
    \begin{tabular}{lcc}
        \hline
        \hline
        Noise Parameter &  Direct-RF SDR Prototype\\
        \hline
        \hline
        $T_n^{Tx-Rx}$ & $3500$~K\\
        $S_{\phi}(f_{\mathrm{mod}})$ & $-105$~dBc/Hz\\
        $S_{\gamma}(f_{\mathrm{mod}})$ & $-129$~dBc/Hz\\
        \hline 
        \hline
    \end{tabular}
\end{table}

While these values allow quantifying the performance of the SDR system, it is not possible to estimate the impact on the readout noise without considering its interaction with the cryogenic microwave chain. The impact of the complete readout system will be evaluated below.

\subsection{Cryo-Loopback Characterization}

\begin{figure}[!t]
\centering
\includegraphics[width=0.48\textwidth]{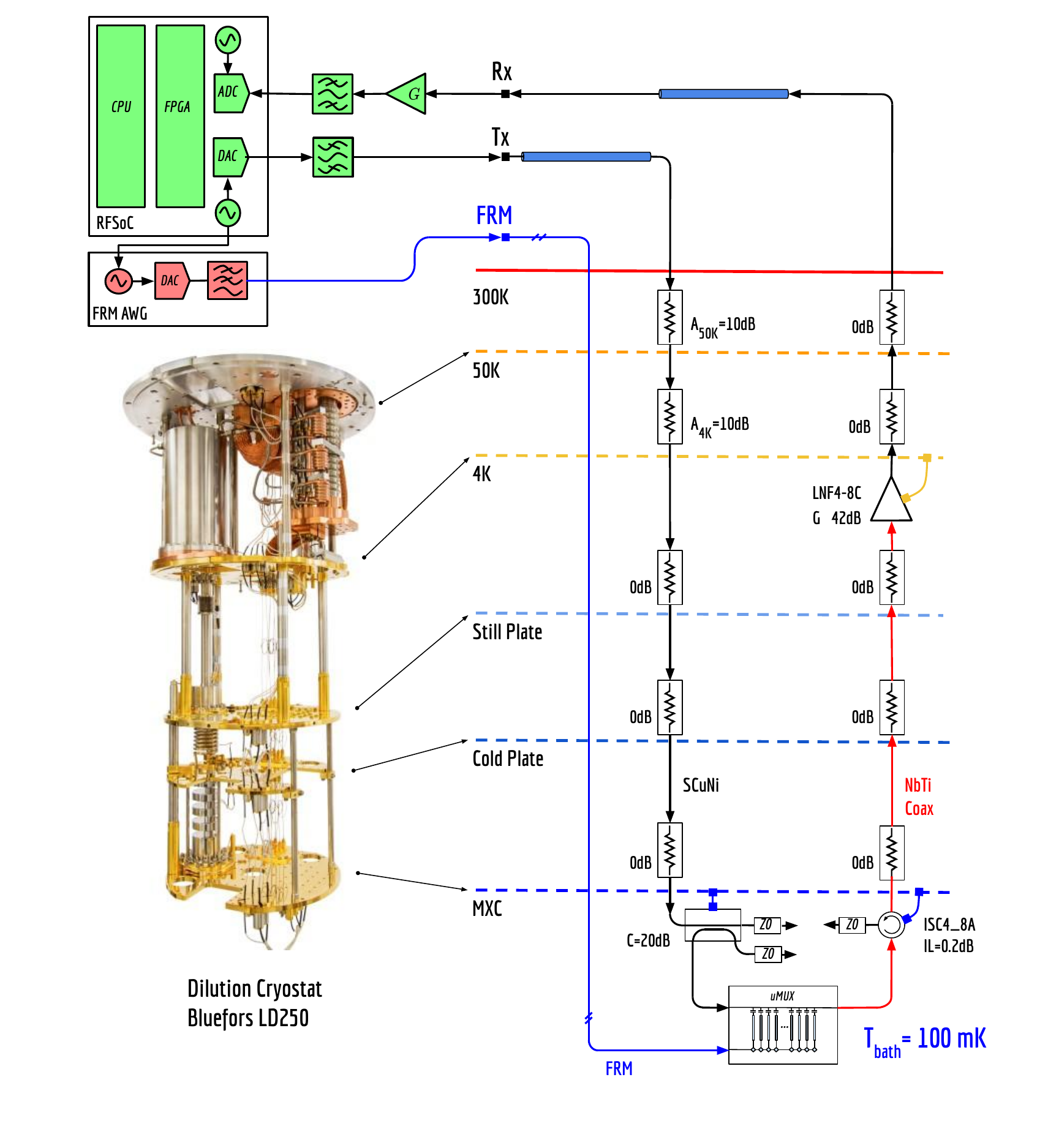}
\caption{Experimental set up for the \textmu MUX readout demonstration using the Direct-RF SDR prototype system. The AWG is used in order to generate the flux modulation.}
\label{fig:cryosetup}
\end{figure}

\noindent As shown in Figure~\ref{fig:cryosetup}, the Direct-RF SDR prototype was connected to the cryogenic microwave chain alongside our nine-channel \textmu MUX prototype. The base operating temperature was set to $T_{base}=100$~mK. This configuration is referred to as Cryo-Loopback. Since the system does not yet have flux-ramp modulation capabilities~\cite{Mates2012}, a \textit{Picotest G5100A} Arbitrary Waveform Generator (AWG) was used instead. The AWG was locked to the \textit{CLK104} frequency reference output and connected to the \textmu MUX modulation line through a twisted-pair cable. A set of attenuators and filters was included in order to optimize the AWG noise performance.

Phase and amplitude noise measurements were performed again as a function of excitation power, as described in Section~\ref{subsec:rfloop}. However, in this case, the measurement was carried out off-resonance, where the \textmu MUX behaves as a direct Tx–Rx connection at the cryogenic stage. In contrast to the RF-Loopback, the power values were referenced to the input port of the \textmu MUX ($P_{\mathrm{exc}}$) using the nominal attenuation of the cryogenic transmission chain, $A_{Tx}=45$~dB. The range of excitation power used was $[-100,-60]$~dBm. The results of the phase and amplitude noise measurements as a function of excitation power are shown in Figure~\ref{fig:nsd_cryo}. 

From the obtained results, the total system noise temperature $T_{n}^{\mathrm{tot}}$ as well as the multiplicative noise values $S_{\phi}(f_{\mathrm{mod}})$ and $S_{\gamma}(f_{\mathrm{mod}})$ were determined. The total noise temperature $T_{n}^{\mathrm{tot}}$ referred to the output of the \textmu MUX can be calculated using the following equation,

\begin{equation}
    T_{n}^{\mathrm{tot}}=T_n^{Tx}G_{Tx}+T_n^{\mathrm{cryo}}+T_n^{Rx}/G_{Rx},
    \label{eq:tempcalc}
\end{equation}

where $T_n^{Tx}$, $T_n^{cryo}$, and $T_n^{Rx}$ are the equivalent noise temperatures of the transmitter, the cryogenic microwave chain, and the receiver, respectively. $G_{Tx}$ and $G_{Rx}$ are the gains of the transmission and reception paths of the cryogenic microwave chain, respectively. Previously, using a vector network analyzer (VNA) \textit{Keysight PNA-X N5242B}, these values were measured, yielding $G_{Tx}=-45$~dB, $G_{Rx}=32.5$~dB, and $T_n^{\mathrm{cryo}}=4.6$~K. Here, for the noise temperature measurement the cold-source method was implemented~\cite{coldsource}. By combining Equation~\ref{eq:tempcalc} with $T_n^{Tx-Rx}=T_n^{Rx}+T_n^{Tx}$, the contributions of the SDR transmitter and receiver were extracted. As a result, the calculated noise temperature together with the multiplicative noise values are listed in Table~\ref{tab:noisesweepcryo}.

\begin{table}[h]
   \caption{Summary of the additive and multiplicative noise levels for the Direct-RF RFSoC-Based SDR prototype measured in Cryo-loopback and referenced to the \textmu MUX output.}
    \centering
    \begin{tabular}{lcc}
        \hline
        \hline
        Noise Parameter &  Direct-RF SDR Prototype\\
        \hline
        \hline
        $T_n^{tot}$ &  $6.1$~K\\
        $T_n^{cryo}$ &  $4.6$~K\\
        $T_n^{Tx}$ & $0.03$~K\\
        $T_n^{Rx}$ &  $1.47$~K\\
        $S_{\phi}(f_{\mathrm{mod}})$ &  $-102$~dBc/Hz\\
        $S_{\gamma}(f_{\mathrm{mod}})$ &  $-122$~dBc/Hz\\
        \hline
        \hline
    \end{tabular}
    \label{tab:noisesweepcryo}
\end{table}

These results allow us to conclude that the total additive noise temperature ($T_{n}^{\mathrm{tot}}$) is primarily dominated by the cryogenic microwave chain, $T_{n}^{\mathrm{cryo}}$. As expected, the measured value is on the order of the typical values for the cryogenic amplifier used \textit{LNF-LNC4-8C}~\cite{LNC48}. The second most significant contribution arises from the SDR receiver ($T_{n}^{\mathrm{Rx}}$) and can be mitigated increasing the gain of the cryogenic LNA. However, this could reduce the linearity of the chain~\cite{Yu2023-dk}. Regarding the Multiplicative noise, it is still dominated by the phase noise of the sampling clocks $S_{\phi}(\Delta f)$. Although an increase is observed with respect to the RF-Loopback configuration, the degradation is associated with the coherence of the synthesizers due to the coaxial cable delay and the drift/vibrations in the cryogenic microwave system~\cite{Silva2022}.

 \section{\textmu MUX Readout Demonstration}

\noindent Once the noise of the SDR system was characterized, a demonstration of the \textmu MUX readout was carried out. A single \textmu MUX channel was arbitrarily selected, and its multidimensional characterization was performed to determine the optimal readout parameters. Using the SDR system and the AWG, the transmission scattering parameter $S_{21}$ was measured as a function of the modulation flux $\Phi_{\mathrm{mod}}$, excitation frequency $f_{\mathrm{exc}}$, and excitation power $P_{\mathrm{exc}}$. From the maximum and minimum resonance frequencies, $f_r^{\mathrm{max}}(f_{\mathrm{exc}}, P_{\mathrm{exc}},\Phi_{\mathrm{mod}})$ and $f_r^{\mathrm{min}}(f_{\mathrm{exc}}, P_{\mathrm{exc}},\Phi_{\mathrm{mod}})$, the optimal trajectories $(f_{\mathrm{exc}}, P_{\mathrm{exc}})$ for minimum noise were determined as described in~\cite{Garcia2024}.

For this readout demonstration, first, the AWG was configured to generate a sawtooth signal with frequency $f_{\mathrm{ramp}} = 1 \cdot 10^{9}/2^{19} \approx 1907$~Hz and an amplitude equivalent to $n_{\Phi_0}=4$. This corresponds to a flux modulation frequency of $f_{\mathrm{mod}} = f_{\mathrm{ramp}} \cdot n_{\Phi_0} = 7629$~Hz. Then, for each readout parameter configuration $(f_{\mathrm{exc}}, P_{\mathrm{exc}})$, an IQ time trace $x(t)=x_I(t)+j\,x_Q(t)$ was acquired with a total of $N=2^{25}$ samples. Subsequently, the resonator phase $\theta(t)$ and amplitude $\gamma(t)$ traces were calculated off-line and flux-ramp demodulated~\cite{Mates2012}. Finally, the flux spectral density was estimated and the white noise level determined. In order to obtain a better estimate of the noise level, $N=20$ spectra were averaged. Particularly for this case, no periods were discarded ($n_{\mathrm{disc}}=0$) during flux-ramp demodulation~\cite{Garcia2024}.

\begin{figure*}[t]
\centering
    \centering
    \captionsetup[subfigure]{
        singlelinecheck = false,
        justification=raggedright,
        captionskip=0mm,
        position=top
    }
    \subfloat[\label{fig:noise_fexc}]{{\includegraphics[width=0.48\textwidth]{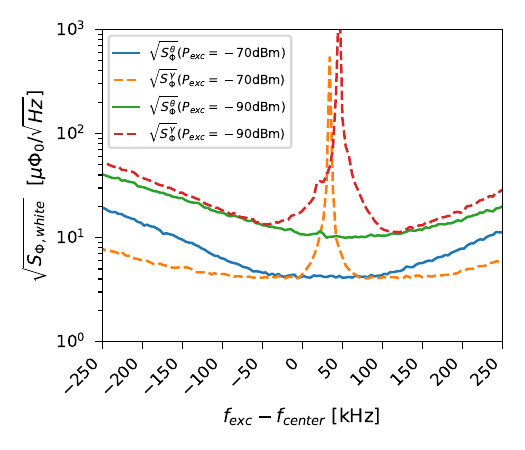} }}%
    \subfloat[\label{fig:noise_pexc}]{{\includegraphics[width=0.48\textwidth]{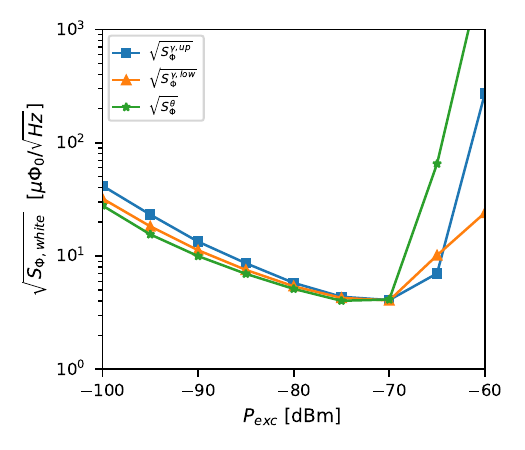} }}%
    \label{fig:example}
\caption{(a) Flux-ramp demodulated white flux noise density $\sqrt{S_{\Phi,\mathrm{white}}}$ for both resonator phase $\theta$ and amplitude $\gamma$ demodulation domains measured for two different excitation powers $P_{\mathrm{exc}}$ using the Direct-RF SDR prototype. Here, $f_{\mathrm{center}}=6.44940$~GHz . (b) Flux-ramp demodulated white flux noise density $\sqrt{S_{\Phi,\mathrm{white}}}$ for both demodulation domains measured under the minimum noise trajectories ($P_{\mathrm{exc}}$,$f_{\mathrm{exc}}$) using the Direct-RF SDR prototype.}
\end{figure*}

The single-channel flux-ramp demodulated white flux noise $\sqrt{S_{\Phi,white}}$ was measured in both demodulation domains (amplitude $\gamma$ and resonator phase $\theta$) as a function of excitation frequency $f_{\mathrm{exc}}$ and power $P_{\mathrm{exc}}$. Figure~\ref{fig:noise_fexc} shows the demodulated flux noise in both demodulation domains as a function of $f_{\mathrm{exc}}$ for two different excitation powers $P_{\mathrm{exc}}$. These results are consistent with previous simulations~\cite{Garcia2024} and highlight the importance of selecting the appropriate readout domain and optimal parameters to achieve the minimum readout noise. As can be seen in the Figure ~\ref{fig:noise_fexc}, for low powers the optimum demodulation domain is the resonator phase domain $\theta$ and the position of the tone is close to the center frequency $f_{\mathrm{center}}=6.44940$~GHz. This center frequency was arbitrarily selected in order to reference the measurements with respect to that value. In contrast, and as explained in the literature~\cite{Garcia2024}, at high excitation powers both domains become comparable, since only the phase demodulation domain ($\theta$) is affected by the increased phase multiplicative noise.

Furthermore, Figure~\ref{fig:noise_pexc} shows the demodulated flux noise for both demodulation domains along the previously determined optimal trajectories $(f_{\mathrm{exc}}, P_{\mathrm{exc}})$. As can be seen, the minimum noise value is $\sqrt{S_{\Phi}^{\theta}}=\sqrt{S_{\Phi}^{\gamma}} \approx 4$~$\mu \Phi_0/\sqrt{\mathrm{Hz}}$ in both demodulation domains and reached at an excitation power of $P_{\mathrm{exc}} \approx -73$~dBm. This value is consistent with previous simulations and lies few decibels below the \textmu MUX insensitivity point at approximately $P_{\mathrm{exc}} \approx -63$~dBm.

Although the measured values are comparable to those reported in the literature~\cite{Bennett2019,Dober2017,Malnou2023}, future work will address the optimization of the receiver chain noise temperature and the improvement of the electromagnetic shielding.

\section{Conclusion}

\noindent We presented the development of a Direct-RF Software-Defined Radio (SDR) system based on RFSoC devices. This design reuses previous firmware developments, inheriting their functionalities while providing a highly integrated system.

The system was characterized in transmission, demonstrating its ability to generate signals across the $4$–$8$~GHz band, which is typical for \textmu MUX readout systems. The output power levels and spurious-free dynamic range are comparable to those achieved with traditional systems that rely on frequency-conversion boards. Although Direct-RF SDR systems impose stringent filtering requirements, they eliminate the need for IQ and LO-leakage calibration. It has been shown that the analog performance of the SDR system, combined with the cryogenic RF chain, is comparable to that of conventional systems. Additive noise was found to be primarily dominated by the cryogenic microwave chain and the SDR receiver, while multiplicative noise was negligible.   The minimum demodulated white flux noise $\sqrt{S_{\Phi}}$ obtained at the optimal readout parameters was $\approx 4 \mu\Phi_0/\sqrt{Hz}$ for both the amplitude $\gamma$ and resonator phase $\theta$ demodulation domains. These results are comparable to those reported in the literature.

We conclude that the development of Direct-RF SDR systems is in line with the new technologies available on the market and represents the next generation of radio systems for the frequency-multiplexed readout of superconducting devices.

\balance

\section*{Acknowledgments}
M. E. García Redondo is supported by the Comisión Nacional de Energía Atómica (CNEA) as well as for the Helmholtz International Research School in Astroparticles and Enabling Technologies (HIRSAP). M. E. García Redondo also acknowledges the support of the Karlsruhe School of Elementary and Astroparticle Physics: Science and Technology (KSETA).


\end{document}